# The calibration of the Sudbury Neutrino Observatory using uniformly distributed radioactive sources.

K. Boudjemline[i], B. Cai[i], B.T. Cleveland[d,j], H.C. Evans[i], G.T. Ewan[i],
J. Farine[d], R.J. Ford[d,j], E. Guillian[i], A.L. Hallin[a,i], E.D. Hallman[d],
C. Howard[a,i], P. Jagam[c], N.A. Jelley[g], K.J. Keeter[i], J.R. Klein[k,h], C. Kraus[i],
C.B. Krauss[a,i], R. Lange[b], I.T. Lawson[c,j], J.R. Leslie[i], J.C. Loach[g,e],
A.B. McDonald[i], G. McGregor[g], A.J. Noble[i], H.M. O'Keeffe[i,g],
S.J.M. Peeters[g,*], A.W.P. Poon[e], S.D. Reitzner[c], K. Rielage[f,l],
R.G.H. Robertson[l], V.L. Rusu[k], S.R. Seibert[f,k], P. Skensved[i], M.J. Thomson[i]

[a]*Dept. of Physics, University of Alberta, Edmonton, AB T6G 2G7, Canada*
[b]*Chemistry Dept., Brookhaven National Laboratory, Upton, NY 11973-5000, USA*
[c]*Physics Dept., University of Guelph, Guelph, ON N1G 2W1, Canada*
[d]*Dept. of Physics, Laurentian University, Sudbury, ON P3E 2C6, Canada*
[e]*Institute for Nuclear and Particle Astrophysics and Nuclear Science Division, Lawrence Berkeley National Laboratory, Berkeley, CA 94720, USA*
[f]*Los Alamos National Laboratory, Los Alamos, NM 87545, USA*
[g]*Dept. of Physics, University of Oxford, Denys Wilkinson Building, Keble Road, Oxford, OX1 3RH, UK*
[h]*Department of Physics and Astronomy, University of Pennsylvania, Philadelphia, 18 PA 19104-6396, USA*
[i]*Dept. of Physics, Queen's University, Kingston, ON K7L 3N6, Canada*
[j]*SNOLAB, Sudbury, ON P3Y 1M3, Canada*
[k]*Dept. of Physics, University of Texas at Austin, Austin, TX 78712-0264, USA*
[l]*Center for Experimental Nuclear Physics and Astrophysics, and Dept. of Physics, University of Washington, Seattle, WA 98195, USA*

**Abstract**

The production and analysis of distributed sources of $^{24}$Na and $^{222}$Rn in the Sudbury Neutrino Observatory (SNO) are described. These unique sources provided accurate calibrations of the response to neutrons, produced through photodisintegration of the deuterons in the heavy water target, and to low energy betas and gammas. The application of these sources in determining the neutron detection efficiency and response of the $^3$He proportional counter array, and the characteristics of background Cherenkov light from trace amounts of natural radioactivity is described.

*Keywords:* radioactive calibration sources, $^{24}$Na, $^{222}$Rn, solar neutrino, SNO
*PACS:* 29.50.-n, 26.65.+t, 81.20.Ym

*Corresponding Author: University of Sussex, Physics and Astronomy, 4A5 Pevensey II, Falmer, Brighton, BN1 9QH, UK. Tel: (+44) (0)1273 678128 Email: S.J.M.Peeters@sussex.ac.uk

# 1. Introduction

## 1.1. Overview of SNO

The Sudbury Neutrino Observatory (SNO) [1] was a 1 ktonne heavy water Cherenkov detector located at a depth of 2092 m (5890 ± 94 m of water equivalent) in Vale INCO's Creighton mine near Sudbury, Ontario in Canada. The detector (see Fig. 1) consisted of a 5.5 cm thick, 12 m diameter acrylic vessel (AV), holding the 1000 tonnes of ultra-pure $D_2O$ target, surrounded by 7 ktonnes of ultra-pure $H_2O$ shielding. The AV was surrounded by a 17.8 m diameter geodesic sphere (PSUP), holding 9456 inward-looking and 91 outward-looking 20 cm photomultiplier tubes (PMTs). Each PMT had a light concentrator that increased the light collection efficiency. With this, the effective PMT coverage was ∼54%. The volume outside the geodesic support structure acted both as a cosmic-ray veto system and as a shield from naturally occurring radioactivity in the surrounding rock and construction materials.

Neutrinos from $^8B$ decays in the Sun interact with $D_2O$ in the following three ways:

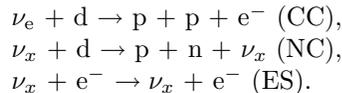

$\nu_e + d \to p + p + e^-$ (CC),
$\nu_x + d \to p + n + \nu_x$ (NC),
$\nu_x + e^- \to \nu_x + e^-$ (ES).

where $x = e, \mu, \tau$.

The charged current (CC) reaction is sensitive only to electron neutrinos, while the neutral current (NC) reaction is sensitive to all active neutrino flavours above the energy threshold of 2.225 MeV. The elastic scattering (ES) reaction is sensitive to all flavours as well, but with reduced sensitivity to $\nu_\mu$ and $\nu_\tau$. The CC and ES reactions were observed by detecting the Cherenkov light produced by the electrons in the final state of the reaction. The NC reaction was observed through the detection of the neutron in the final state, using three different methods. These three methods distinguish the three operational phases of the SNO experiment.

In the first phase of SNO (phase I), from November 1999 to May 2001, during which the AV contained only pure $D_2O$, neutrons from the NC reaction were detected by the Cherenkov light produced by the interaction of the 6.257 MeV $\gamma$-ray emitted following neutron capture on deuterium. SNO observed a significant difference between the equivalent neutrino fluxes measured with the CC and NC reactions. Results from the first phase provided the first direct evidence for neutrino flavour transformation [2, 3].

In the second phase of SNO (phase II), from June 2001 to October 2003, two tonnes of ultra-pure salt (NaCl) were added to the heavy water, which increased the neutron capture efficiency and the Cherenkov light associated with each event. The neutrons captured predominantly on $^{35}Cl$, nearly always leading to the emission of multiple $\gamma$-rays with a total energy of about 8.6 MeV. The resulting Cherenkov light patterns of PMT hits were generally more isotropic than those from the relativistic electrons produced in the CC and ES reactions.

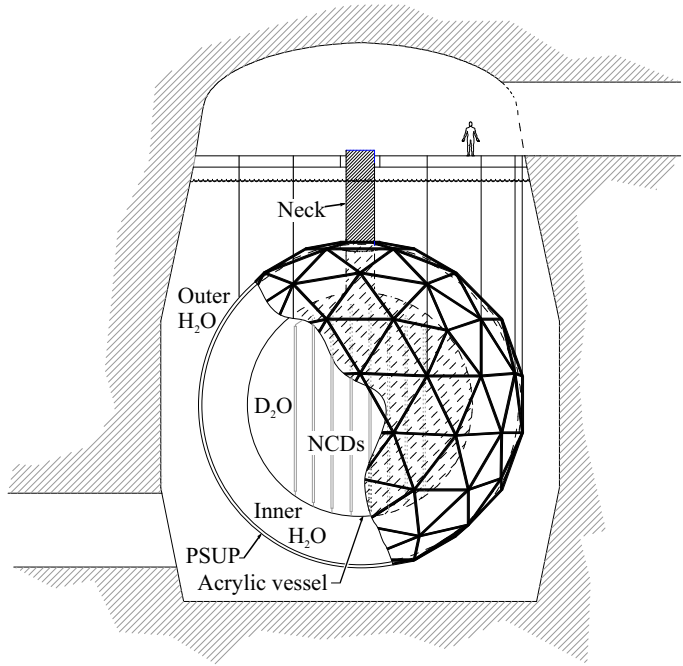

Figure 1: The layout of the SNO detector. The $D_2O$ was contained within an acrylic vessel. This volume could be accessed via the neck. The Neutral Current Detector (NCD) array was deployed in the third phase of the experiment. The acrylic vessel was surrounded by water for shielding the radioactivity from the rock (outer $H_2O$) and the photomultiplier tubes (inner $H_2O$). The Photomultiplier SUPport structure (PSUP) held the photomultiplier tubes and light concentrators.

This difference in isotropy allowed good statistical separation of the event types and a precise NC measurement of the $^8$B neutrino flux to be made, independent of assumptions about the CC and ES energy spectra. The results from this phase provided further and definitive evidence of neutrino flavour change, and the precise determination of the total active solar neutrino flux was found to be in excellent agreement with the predictions from solar models [4, 5].

In the third and final phase of SNO (phase III), from November 2004 to November 2006, the neutrons produced by the NC reaction were observed by a $^3$He-filled, ultra-low radioactivity proportional counter array, called the Neutral Current Detection (NCD) array [6]. This array was deployed in the heavy water after the salt was removed in the summer of 2003 and changed the method of neutron identification from that used in the first two phases.

The NCD and PMT arrays were also effectively independent detectors with different associated systematic uncertainties. Thus, using the PMT array to detect CC events and the NCD array to detect most NC events further reduced the correlation between the measurements of the NC and CC fluxes, providing an independent check on the first two phases. The NC and CC flux measurements determined in phase III [7] were in excellent agreement with the results from previous phases.

*1.2. Calibration of the SNO detector*

The results from SNO depended on very accurate calibration of both the energy and position of an event and of the isotropy of the light produced by an event. Through the CC, ES and NC interactions, the solar neutrinos produced electrons and neutrons distributed uniformly throughout the heavy water volume. The response of the SNO detector was principally measured by using a number of contained, point-like sources [3] placed at known positions throughout the heavy water or in the surrounding light water. In particular, a $^{16}$N source [8] was used to calibrate the energy and position reconstruction by measuring the response to ∼5 MeV kinetic energy electrons produced by Compton scattering of 6.129 MeV $\gamma$-rays from the decay of $^{16}$N. For the response to neutrons, $^{252}$Cf and AmBe sources were deployed.

The energy of an event was calibrated in terms of an equivalent average electron kinetic energy. The observed energy based on this calibration is called the effective kinetic energy. Analytic models could be used as well as Monte Carlo simulations to interpolate between and extrapolate from source positions, while the Monte Carlo was used to correct for the small effects of absorption in the source containers and deployment apparatus. The Monte Carlo program that was used in all the studies in this paper was the SNOMAN code [1].

The NCD array deployed in the third and final phase of SNO, however, caused the detector response to deviate significantly from spherical symmetry (see Fig. 2). This made analytic interpolation and extrapolation from point source measurements difficult and we relied more heavily on the Monte Carlo to characterize the relative response of the NCD array. This process was made more difficult than in previous phases because the discrete location of the NCDs made source position uncertainties much more significant. To overcome this limitation a uniformly distributed source of neutrons was required.

The lower energy threshold for the detection of solar neutrinos was limited by the background due to Cherenkov light from trace amounts of natural radioactivity in the detector. Moreover, this activity led to a background to the Neutral Current signal: a deuteron can be dissociated by a $\gamma$-ray above its binding energy of 2.225 MeV. The $\beta-\gamma$ decays of $^{208}$Tl and $^{214}$Bi from the naturally occurring $^{232}$Th and $^{238}$U chains emit $\gamma$-rays that are above this binding energy. Therefore, the measurement of the levels of radioactivity inside the detector was crucial for a correct measurement of the total $^8$B flux.

Three methods, classified as ex situ techniques, were developed to assay precursor radioisotopes of $^{208}$Tl and $^{214}$Bi in the D$_2$O and the H$_2$O [9, 10, 11, 12]. Another method of determining the amounts of $^{208}$Tl and $^{214}$Bi directly from the data was developed. This in situ technique, analysed events that occurred in a low energy window (4.0−4.5 MeV effective kinetic energy), in which the selected events were dominated by $\beta-\gamma$ decays from these nuclei [4]. To determine the response to $^{208}$Tl and $^{214}$Bi decays, distributed low-energy $\beta-\gamma$ sources were required to avoid the effect of attenuation of the betas in contained sources of Th and U.

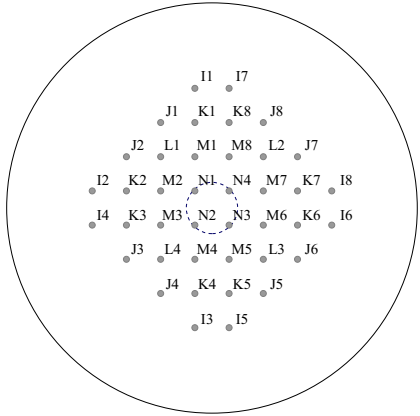

Figure 2: The positions of the NCD strings projected onto the plane of the AV equator (looking downwards). The letters denote strings of the same length and distance from the centre of the AV. The outer circle indicates the AV equator and the inner circle indicates the position of the neck of the AV. (North is upwards.)

This paper describes the development of two novel distributed and uncontained sources. These were: $^{24}$Na, which provided a calibration of the response of the SNO detector to low energy $\beta-\gamma$ decays and, through the photodisintegration of deuterons in the heavy water detector, to neutrons; and $^{222}$Rn, which is part of the $^{238}$U chain or a direct $\beta-\gamma$ contaminant from the ingress of traces of mine air.

First, the methods used to prepare the sources and measure their strengths are explained. Then, the techniques that were used to mix and measure the distributions of $^{24}$Na and $^{222}$Rn throughout the D$_2$O are described. Finally, the results for the neutron detection efficiency and response in phase III and for the verification of the Monte Carlo simulation of radioactive backgrounds in phase II and III are presented.

## 2. Distributed calibration sources of $^{222}$Rn and $^{24}$Na in SNO

The nucleus $^{24}$Na provided neutrons through photodisintegrating deuterons in the heavy water: $^{24}$Na ($Q_{\beta^-} = 5.516$ MeV) $\beta$ decays nearly always to the second excited state of $^{24}$Mg at 4.123 MeV, with an end-point energy of 1.393 MeV. The 4.123 MeV state then $\gamma$ decays to the 1.369 MeV first excited state emitting a 2.754 MeV $\gamma$-ray. On average, one in every 380 $\gamma$-rays of this energy can photodisintegrate the deuteron (binding energy 2.225 MeV) emitting a neutron.

Table 1: List of controlled spikes of $^{24}$Na and $^{222}$Rn

| Spike | ♯ | Phase | Region | Method | kBq |
|---|---|---|---|---|---|
| $^{24}$Na | 1 | II | D$_2$O | Th | $\sim 0.006$ |
| $^{24}$Na | 2 | II | D$_2$O | Th | $\sim 0.007$ |
| $^{222}$Rn | 3 | II | D$_2$O | Air | $\sim 0.08$ |
| $^{222}$Rn | 4 | II | H$_2$O | Air | $\sim 0.8$ |
| $^{24}$Na | 5 | III | D$_2$O | Reactor | $\sim 0.48$ |
| $^{222}$Rn | 6 | III | D$_2$O | Air | $\sim 0.20$ |
| $^{24}$Na | 7 | III | D$_2$O | Reactor | $\sim 0.32$ |
| $^{222}$Rn | 8 | III | H$_2$O | Ra | $\sim 50$ |

Provided the $^{24}$Na nuclei could be distributed uniformly throughout the heavy water, a near-uniform distribution of neutrons could be generated to measure the neutron efficiency of the SNO detector. The decay of $^{24}$Na also provided a calibration of the response of SNO to low energy $\beta - \gamma$ decays.

The 14.96 hour half-life of $^{24}$Na was very suitable as it gave sufficient time for mixing but was not so long as to disrupt data taking significantly: the activity could be thoroughly mixed within a few days. After this there were still enough neutrons produced for an accurate calibration to be obtained before the $^{24}$Na had decayed away sufficiently for neutrino data taking to resume, about 10 days after it was introduced.

The nucleus $^{222}$Rn is part of the $^{238}$U chain and decays with a 3.82 day half-life via $^{218}$Po and $^{214}$Pb to $^{214}$Bi. The nucleus $^{214}$Bi decays predominantly to $^{214}$Po with the emission ($\sim 2.5\%$ of the time) of $\gamma$-rays with energies greater than 2.225 MeV that could photodisintegrate deuterons. The greater half-life of $^{222}$Rn compared with $^{24}$Na required a longer time before data taking could be resumed after a $^{222}$Rn spike was introduced.

The $\beta$ and $\gamma$ rays from $^{214}$Bi dominated the low energy background Cherenkov light in SNO below 4.5 MeV effective kinetic energy. Therefore, a radon spike could provide a test of the Monte Carlo description of the detector response to radon and of the yield of neutrons through deuteron photodisintegration.

A number of controlled spikes of $^{24}$Na and $^{222}$Rn were made during the course of the SNO experiment and these are listed in Table 1. The method refers to the technique for production of the source:

Th - 2.614 MeV $\gamma$-rays from $^{232}$Th source generated neutrons through photodisintegration of deuterons. Subsequent capture of these neutrons by $^{23}$Na in the heavy water brine produced $^{24}$Na;

Air - $^{222}$Rn extracted from mine air by use of a liquid N$_2$ cooled trap;

Reactor - $^{24}$Na produced through thermal reactor neutron activation of a brine solution;

Ra - $^{222}$Rn produced from the decay of a $^{226}$Ra source.

## 2.1. The $^{222}$Rn sources extracted from mine air

In phase I of SNO there were two periods in which accidental periods of high radon concentrations (spikes) in both the $D_2O$ and the $H_2O$ occurred. These provided a useful verification of the Monte Carlo estimates for the background Cherenkov light contributions to the data [3].

For a more accurate assessment in phase II, a calibrated radon spike for the $D_2O$ (spike ♯3) was prepared by using a mobile radon-assay board, which had a high-efficiency cryogenic absorption trap. The board was used to trap and concentrate natural Rn from the air (the underground $^{222}$Rn activity was $\sim$100 Bq/m$^3$) and transfer it to 15 cm$^3$ Lucas cells for $\alpha$-scintillation counting. The air was first passed through two cartridges containing NaOH pellets, which absorbed $CO_2$ in the air and a trap held at -85 C which removed $H_2O$ vapor. The Rn was collected in a trap, transferred to the Lucas cells, and then dissolved in a four liter sample of $D_2O$ that had been thoroughly degassed. The aqueous (saline) Rn source was then transferred quickly (within one minute) into a four liter injection chamber so as to minimize the amount of Rn coming out of solution.

There was a total of $81 \pm 4$ Bq of $^{222}$Rn dissolved and this solution was injected directly into the heavy water in the acrylic vessel (spike ♯3). The concentration of Rn was equivalent[1] to $6.5 \times 10^{-12}$ gU/gD$_2$O, which was close to 800 times the measured concentration in the $D_2O$ of $8.3 \times 10^{-15}$ gU/gD$_2$O [4] during solar neutrino data taking. Five samples of the solution were taken during injection to determine the amount of radon entering the detector. The samples were counted via an organic liquid scintillation coincidence counting technique that had been developed by the HTiO Ra assay group [9, 12]. The liquid scintillation technique used to determine the $^{222}$Rn activities was calibrated with a $^{226}$Ra standard source before, during and after counting the samples.

## 2.2. $^{222}$Rn calibration source produced from a $^{226}$Ra source

The same method was used for producing the radon used in the $H_2O$ spike in phase II (spike ♯4) but, in this case, the radon was introduced by connecting the Lucas cells to a membrane re-gasser and purging the cells with nitrogen to ensure efficient gas transfer into the $H_2O$. Two injections, each of $\sim 500$ Bq, were made five days apart. However, the resulting maximum source strength of $\sim$ 800 Bq proved too low to provide sufficient statistics to study the response of the detector to activity in the $H_2O$, but was used to calibrate the in situ technique in the $H_2O$. The same technique was also used in phase III to provide a uniformly distributed $^{222}$Rn source of known strength ($194 \pm 15$ Bq) to calibrate the radon assay technique [11]. For this spike ♯6, the $^{222}$Rn was introduced by connecting

---

[1]The $^{232}$Th and $^{238}$U chains were not in secular equilibrium in the SNO heavy water due to leaching of radium isotopes from trace contamination of materials in contact with the water and from ingress of $^{222}$Rn (half-life of 3.8 days). However, to facilitate comparisons between different assay methods, concentrations are quoted in terms of the equivalent amounts of $^{232}$Th and $^{238}$U in a chain in equilibrium.

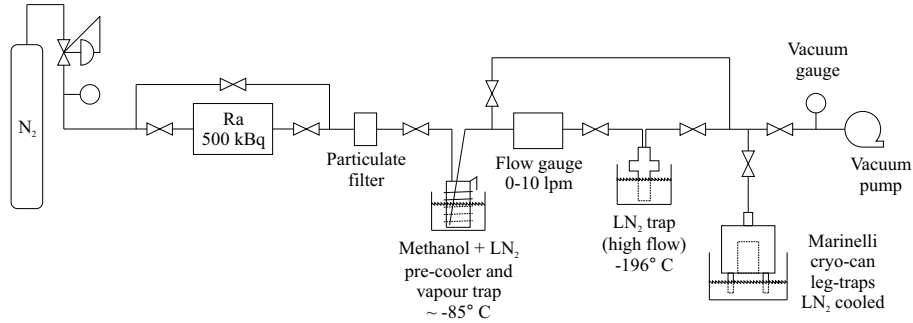

Figure 3: Rn production system using a Pylon Ra source

the Marinelli can containing the radon to the membrane re-gasser for the heavy water.

Radon in the $H_2O$ could give rise to events that mis-reconstructed in the $D_2O$, and also to neutrons from $\gamma$-rays with energies greater than 2.225 MeV that penetrated into the $D_2O$. To determine the number of these events required a source strength of around 20 kBq, which was far larger than the strengths that could feasibly be extracted from mine air (of $\sim$ 100 Bq). So, instead, a commercial Ra source (Pylon RN-1025-500[2]) which produces Rn gas was used as the input to the cryo-trap on the mobile radon-assay board (see Fig. 3). Using the Pylon Ra source with a nitrogen carrier-gas greatly improved and simplified the Rn collection, as the source was strong and there was no interference with $CO_2$ in the cryo-trapping.

The radon gas was transferred from the cryo-trap to a custom designed stainless steel can with a Marinelli geometry to give good heat transfer and because radon easily diffuses through plastic. The Marinelli can also had legs, which were placed into a liquid nitrogen bath and acted as cold-fingers to freeze out the radon when the cryo-trap was warmed up. The Marinelli can was placed on the SNO germanium detector (see Section 2.4.1 below) and the radon activity was determined to be $50 \pm 2.4$ kBq. This was injected as spike ♯8.

It was very important to ensure that no radium escaped from the Pylon source to avoid contamination of the SNO water systems, where it could have become a persistent background. This concern was addressed by flowing the $N_2$ carrier gas used through the source and immediately bubbling it through water without any further filtering or gas transfer stages. The water was then removed and purged with dry nitrogen to remove the dissolved radon, before being tested for any dissolved radium.

A disk consisting of a manganese oxide ($MnO_x$) coating on a nylon substrate was used to extract any radium in the solution (at previously demonstrated

---

[2]Pylon Electronics Inc. Instrument Manufacturing Division, 147 Colonnade Road, Ottawa, Ontario K2E 7L9 Canada

efficiencies of more than 80%) through the well-known ability of $\text{MnO}_x$ surfaces to efficiently trap radium from aqueous solutions [10].

Any radium is absorbed sufficiently close to the disk surface that $\alpha$-particle energy spectra are well-defined and the presence of radium can be quantitatively determined from its characteristic energy peaks. The disk was first counted to determine any background and then placed for six hours in the solution, which was gently stirred. It was then dried and re-counted. The absence of any radium signal in the disk spectrum led to a maximum possible radium activity in the $\text{H}_2\text{O}$ solution of 0.4 mBq.

This activity corresponded to an upper limit of about $1.9 \times 10^{-17}$ gU/g$\text{H}_2$O for an equivalent concentration of uranium in the region between the acrylic vessel and the PMT support structure. The concentration of radium using HTiO and liquid scintillator counting [9, 12] gave a similar limit of $3 \times 10^{-17}$ gU/g$\text{H}_2$O. For comparison, the actual level of $^{238}$U in this region during phase III was measured [13] to be $3.5 \times 10^{-13}$ gU/g$\text{H}_2$O.

## 2.3. The $^{24}$Na sources in phase II

In phase II of SNO, there were 2 tonnes of NaCl dissolved in the heavy water. A thorium source placed in the heavy water emitted 2.614 MeV $\gamma$-rays from the decay of the progeny nucleus $^{208}$Tl, which could photodistintegrate deuterons producing neutrons. The neutrons generally captured on $^{35}$Cl nuclei, but also occasionally on $^{23}$Na nuclei to produce $^{24}$Na.

The first $^{24}$Na spike (spike ♯1) was made by deploying a contained (330 kBq) Th source in the centre of the acrylic vessel for 20.25 hours. The Th source was then removed and data-taking started. From the probability that 2.614 MeV $\gamma$-rays would photodistintegrate a deuteron (0.22% [14]) and that neutrons produced in this way would activate $^{23}$Na (1.4%), it was estimated that there were $4.7 \times 10^5$ ($\pm 8.4\%$) $^{24}$Na atoms at the start of data taking.

The measured spectrum of the number (NHIT) of PMTs per event with a signal[3] is shown in Fig. 4. Events with NHIT $\leq 40$ are dominated by the Cherenkov light from $\beta$-$\gamma$ decays, while events with NHIT $\geq 50$ are predominantly from the decay of $^{36}$Cl following neutron capture on $^{35}$Cl.

There is good agreement with the spectrum expected from a Monte Carlo simulation, with the predicted and measured number of events in the range $30 \leq \text{NHIT} \leq 40$ ($\sim$3.5-4.5 MeV effective kinetic energy) agreeing within the uncertainty of 9% [15].

A second $^{24}$Na spike (spike ♯2) in phase II was also made with the same technique. This spike was used to study the water flow when circulation was on, which is described in Section 3.1.

---

[3]At the time of this calibration, 50 NHIT approximately corresponded to the energy deposited by a 6 MeV kinetic energy electron.

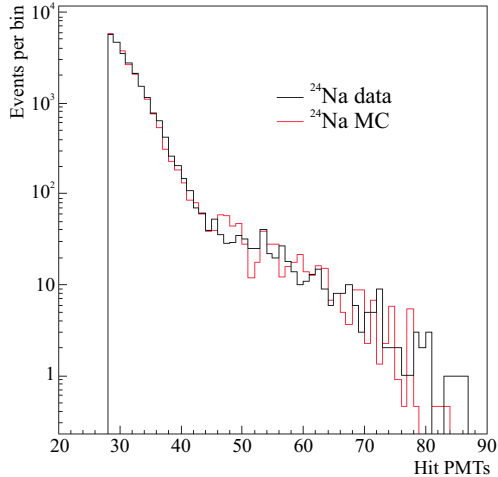

Figure 4: The measured spectrum of the number (NHIT) of PMTs per event observed in SNO after $^{24}$Na-activation using a Th source (spike ♯1), compared to a Monte Carlo simulation. Events with NHIT $\leq 40$ are dominated by the Cherenkov light from $\beta$-$\gamma$ decays, while events with NHIT $\geq 50$ are predominantly from the decay of $^{36}$Cl following neutron capture on $^{35}$Cl.

## 2.4. The $^{24}$Na sources in phase III

In phase III, a 10% brine solution, which was taken as a sample when the NaCl in heavy water was purified before the beginning of the second phase of SNO, provided very pure $^{23}$Na to activate. Only a total amount of 0.45 g of activated salt had to be added, which was a negligible addition to the heavy water. Neutron activation of NaCl also creates $^{36}$Cl and $^{38}$Cl, with half-lives of $3 \times 10^5$ years and 37 minutes, respectively. These were not a concern as $^{36}$Cl does not emit $\gamma$-rays and the activity of $^{38}$Cl dropped to an insignificant level within a few hours after activation. The possible presence of other nuclides that could be produced was carefully evaluated before activation and checked for afterwards by using a germanium detector.

The reactor used for the activation was the SLOWPOKE-2 reactor at the Royal Military College, Kingston, Ontario. About 3.75 ml of brine was activated for 25 minutes with a slow neutron flux of $5 \times 10^{11}$ neutrons cm$^{-2}$ sec$^{-1}$. The activated brine was then taken to SNO and, once underground in the water handling room alongside the SNO cavity, was injected into approximately two liters of pure heavy water. This heavy water was held in a 2.5 liter cylindrical acrylic container under an atmosphere of nitrogen gas, so as to avoid the absorption of any radon from the mine air, and the mixture was thoroughly stirred. The dilution meant a mass of about one kilogram could be injected into the kilotonne of heavy water in the SNO acrylic vessel. This amount was easier to control and could be accurately determined by weighing the acrylic container both before and after the injection [17].

Two $^{24}$Na spikes were made in the D$_2$O: one near the middle of phase III in October 2005 and the other near the end in October 2006. So as to determine the strengths of the spikes very precisely, three detectors were used: a germanium detector, the NCD array and the PMT array. Samples of the diluted activated brine solutions were placed in the centre of the $D_2O$ and on the SNO germanium detector, which was located just outside the SNO underground control room, and their strengths were determined. The NCD and PMT arrays measured the number of neutrons produced per second by the sample. The germanium detector was used to determine the number of 2.754 MeV $\gamma$-rays produced per second by the sample and this was translated to a number of neutrons produced per second using the Monte Carlo simulation to calculate the photodisintegration probability.

*2.4.1. The SNO germanium detector measurement*

The SNO germanium (Ge) detector contains a coaxial high purity p-type crystal. This is mounted in a low background cryostat around which a $N_2$ atmosphere was maintained throughout data taking. The germanium detector was calibrated using a mixed source containing the radioisotopes $^{65}$Zn, $^{85}$Sr, $^{137}$Cs, $^{54}$Mn, and $^{133}$Ba, homogeneously suspended in an epoxy matrix inside a Marinelli beaker, which produced $\gamma$-rays of energies in the range 250-1150 keV. The geometry of the germanium detector (detailing the position of the crystal, the thickness of its dead layer, etc.) was implemented in the SNOMAN Monte Carlo [16], which uses EGS4 [18] to model the propagation of the $\gamma$-rays in the crystal. The Monte Carlo simulation allowed the efficiencies for the detection of $^{24}$Na $\gamma$-rays (allowing for their angular correlations) to be derived from measurements made using calibration sources with different densities and producing $\gamma$-rays of different energies.

The effect of uncertainties in the detector geometry, construction and charge collection was parametrized by allowing the thickness of the dead layer to vary (for a general discussion of this technique see [19]). Fig. 5 shows the resultant germanium detector efficiency curve from the Monte Carlo simulation using the best fit dead layer of 1.094 ± 0.062 mm [16] and the efficiencies measured with the mixed calibration source.

Samples of $^{40}$K as KCl were also considered as calibration sources, both as solid and dissolved in water. KCl sources, however, are not suitable as absolute calibration sources because of the unreliability of manufacturer-measured purities [20]. Nevertheless, they demonstrated the ability of the Monte Carlo to extrapolate correctly between sources made of different materials.

A $\sim$30 ml sample was taken from the 2.5 liter container holding the diluted activated brine and placed in a Marinelli beaker. Approximately one liter of $D_2O$ was added, so the beaker had the same active volume as the mixed source. From the Monte Carlo derived efficiency for detecting the 1.369 MeV $\gamma$-ray, the 2.754 MeV $\gamma$-ray production rate of the sample could be determined. Allowing for the difference in mass of the sample and of the spike and knowing the photodisintegration probability (0.26% for uniformly mixed $^{24}$Na, using the Monte Carlo simulation) of the $D_2O$, the neutron source strength of the injected source could be calculated.

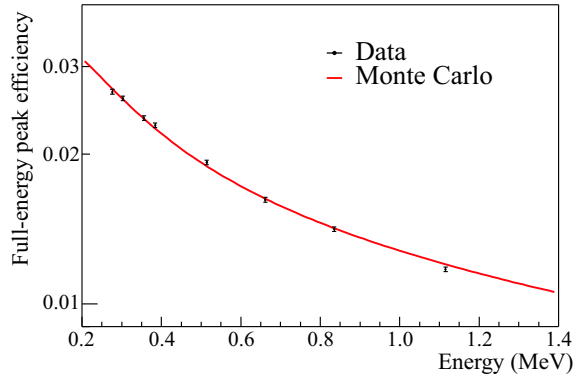

Figure 5: Illustrating the agreement between the Monte Carlo simulated efficiency curve [16] and the measured efficiencies for the SNO germanium detector, using the best fit dead layer of 1.094 mm

*2.4.2. The NCD and PMT array measurements*

The NCD and PMT arrays were also used to measure the neutron detection rate from a sample of the diluted activated brine placed in the centre of the detector. A ∼10 ml $^{24}$Na sample was taken from the 2.5 liter container and placed in a cylindrical Delrin can. To minimize the risk of any leakage, the $^{24}$Na was doubly encapsulated. The can walls containing the $^{24}$Na brine scattered some of the $\gamma$-rays, resulting in a slight reduction in the average $\gamma$-ray energy and, therefore, in the photodisintegration probability. This scattering did not exist for the injected brine and a correction factor of about 1% for this effect was determined using the Monte Carlo simulation. The neutron detection rate was measured as a function of time and the data were consistent with a purely exponential decay.

A determination of the neutron source rate required knowing the neutron detection efficiency of the NCD and PMT arrays. For the NCD array, this was determined by three different methods: by measuring the number of neutrons detected from a well-calibrated $^{252}$Cf source [4] positioned at the centre of the detector, by using the Monte Carlo, and by performing a Time Series Analysis (TSA) [21] of the number of neutrons detected per fission produced by the same source, also at the centre of the detector. Since the neutrons generated by $^{252}$Cf had a quite different radial and energy distribution from that produced by the canned $^{24}$Na, a small Monte Carlo simulation-based correction factor was necessary when inferring the efficiency for a canned $^{24}$Na source. All three methods gave consistent results for the neutron detection efficiency for a central $^{252}$Cf source:

$0.4435 \pm 0.0035$ (NCD);
$0.4496 \pm 0.0037$ (TSA);
$0.4444 \pm 0.0058$ (MC).

The PMT array detected Cherenkov light from neutrons capturing on deuterons producing 6.257 MeV γ-rays (that subsequently Compton scattered) from the centrally positioned $^{252}$Cf and $^{24}$Na sources. The analysis region was selected to maximise the neutron detection rate while excluding background events. This region lay within $200 \leq r \leq 450$ cm, where $r$ is the radial distance from the centre of the acrylic vessel, and $5 \leq T \leq 9.5$ MeV, where $T$ is the effective kinetic energy. The radial cut was used to eliminate γ-rays from the sources. In the $^{24}$Na sources, these were 1.369 MeV and 2.754 MeV γ-rays from the decay of excited $^{24}$Mg (the beta decay final nucleus) and, in the $^{252}$Cf source, there were γ-rays from the de-excitation of the fission fragments. The $^{24}$Na neutron source rate was obtained from a comparison of the capture rates from the $^{252}$Cf and $^{24}$Na sources, allowing for the difference in neutron distributions from the two sources.

*2.4.3. Results for the strengths of the $^{24}$Na spikes*

The results from the three detectors for each of the two $^{24}$Na spikes are summarised in Table 2 and are consistent.

Table 2: Neutron production rates from the $^{24}$Na spikes at the time of complete mixing

| Data Set | Detector | Rate (neutrons s$^{-1}$) |
|---|---|---|
| 2005 | PMT | $1.284 \pm 0.068$ |
|  | Ge | $1.236 \pm 0.029$ |
|  | NCD | $1.240 \pm 0.020$ |
| 2006 | PMT | $0.810 \pm 0.046$ |
|  | Ge | $0.857 \pm 0.020$ |
|  | NCD | $0.838 \pm 0.013$ |

## 3. Mixing of the distributed $^{24}$Na and $^{222}$Rn sources

The light water and the heavy water were separately circulated both to purify and to assay the water. The circulation was also used to mix the $^{24}$Na and $^{222}$Rn spikes. In the acrylic vessel, the heavy water was either removed from the bottom of the AV and injected in the AV neck, or vice versa; at the bottom, the heavy water was directed tangentially to the AV wall, while in the neck, it was directed both tangentially and radially inward. Slight asymmetries in the flow set up some rotational motion as well as bulk flow of the D$_2$O.

The characteristic time $T_f$ for fluid motion to slow down, using a dimensional estimate, is given by $T_f = D^2/\nu$, where $\nu$ is the kinematic viscosity of heavy water ($\sim 10^{-6}$ m$^2$ s$^{-1}$) and $D$ is the extent of the fluid motion. For water flow in the AV, where $D \sim 1$ m, the characterstic time is $T_f \sim 10^6$ s $\sim 10$ days. Flows would be expected, particularly after the water circulation had been on, to persist for days.

The water flow was also predicted to be partly turbulent when the circulation was on. A typical flow rate was $\sim 150$ liters/min so, at the bottom of the AV where the flow in each jet was $\sim 75$ liters/min, the speed of the flow would have been greater than $\sim 0.005$ m s$^{-1}$, where the diameter of the flow tube was less than 0.5 m. The Reynolds number, given by speed×(characteristic length)/(kinematic viscosity) would therefore be greater than $\sim 2500$ in this region. The increase in the half-width of a turbulent jet with axial distance is about 0.18 [22], so turbulent mixing might be expected to occur for about 1-2 metres away from the entry and exit pipes and that, once formed, eddies would persist and move around within the detector. However, within the detector, the main flow would be streamlined.

The distortion to the flow caused by the NCDs was predicted to be small: A dimensional estimate for the velocity boundary layer is $(\nu L/u)^{0.5}$, where $L$ is the characteristic dimension of the NCD and $u$ is the speed of flow. From the flow pattern seen after the first $^{24}$Na spike in the NCD phase (see Fig. 7), flow speeds across the NCDs were estimated to be $u \sim 3 \times 10^{-5}$ m s$^{-1}$. The diameter of the NCDs was $5 \times 10^{-2}$ m, which gives a boundary layer of $\sim 5$ centimetres.

Good mixing was also expected close to the AV. This was because the heat flow from the light water through the acrylic vessel into the heavy water was predicted to give rise to turbulent convective flow, within a thermal boundary layer of a few centimetres thickness, as the Rayleigh number was estimated to be greater than $10^9$.

### 3.1. Results from the $^{24}$Na spikes in phase II

The water flow in the AV was studied after the two $^{24}$Na spikes (spikes ♯2 and ♯3) in the D$_2$O in phase II. When $^{24}$Na (or the progeny of $^{222}$Rn) decayed in heavy water, the emitted Cherenkov light could be detected by the PMT array. This light was used to monitor the distribution of events and to see how the spikes were dispersed. The location of the activity was determined from the relative timing of the photons detected by the PMT array.

The dispersal of the first $^{24}$Na spike (spike ♯1) in phase II, whose preparation was described in Section 2.3 above, was studied from the time the Th source was removed. There was no circulation of the D$_2$O throughout. The distribution of the Cherenkov light produced by the decay of the $^{24}$Na showed the majority of the activity within a radius of 300 cm after 50 hours. There was a slow drift (on the order of 0.5 m/day), mainly upward, of the initial cloud of $^{24}$Na of radius $\sim$200 cm, possibly caused by the removal of the Th source together with small differences in temperature in the heavy water.

The second $^{24}$Na spike in phase II suggested that there was some turbulent mixing when the circulation was turned on. For this spike ♯2, the contained 330 kBq Th source was positioned on the central vertical axis of the acrylic vessel (AV) below the bottom of the neck, 400 cm above the centre of the AV. The Th source was left at this position for 29 hours and then removed. The heavy water was then left undisturbed for an hour. Next, circulation of the D$_2$O was started, taking heavy water out of the bottom and putting it back in the top, and the evolution of the cloud of $^{24}$Na was studied over 80 hours.

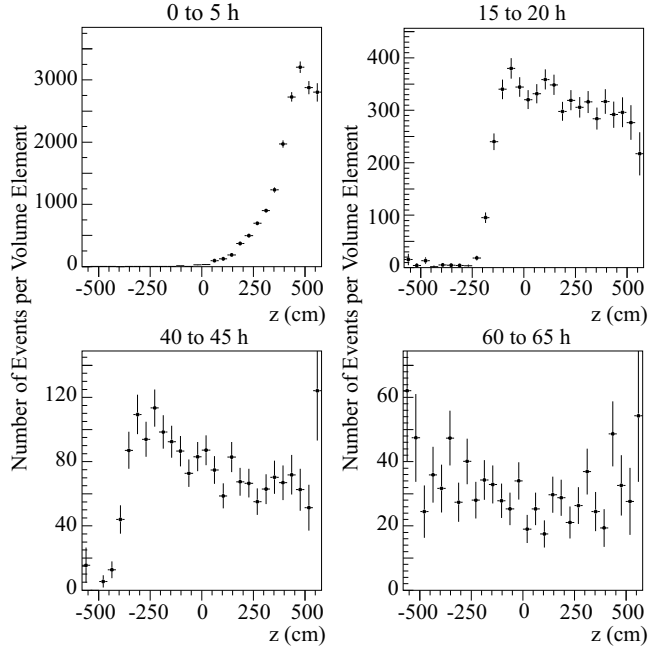

Figure 6: Vertical distribution of $^{24}$Na following the activation run at 400 cm above the center of the AV (spike ♯2). The data is volume weighted, i.e. actual concentrations (in arbitrary units) are plotted. Each plot contains the data of a five-hour time bin and together they cover 0 to 65 hours after the start of circulation.

In the first part of the mixing process, water coming back into the AV mixed with a relatively large fraction from the top of the AV (see Fig. 6). The activity quickly mixed throughout the top half (500 m$^3$) of the vessel within about 15 hours. The flow rate was about 150 liters per minute, so the displaced volume in this time would have been 135 m$^3$. After about 15 hours, an interface formed between unmixed water (containing a very small amount of the $^{24}$Na) and mixed water (containing a large amount of the $^{24}$Na) which moved at a velocity close to that expected from flow measurements. The source can be seen to be relatively well mixed after about 65 hours.

3.2. Methods for injecting the $^{24}$Na and $^{222}$Rn spikes

These results led to the next $^{222}$Rn and $^{24}$Na spikes (♯3 and ♯5) in the heavy water being mixed by first injecting the source in several positions along the central vertical axis of the AV prior to the D$_2$O circulation being turned on. For the October 2005 spike ♯5, the $^{24}$Na began to be distributed throughout the acrylic vessel once the circulation of the heavy water was started.

Fig. 7 shows the distributions of events in the detector following spike ♯5 along a vertical slice enclosing the centre of the vessel. These plots show the event rate in equal volume cubes. Most of the activity was initially at the bottom of the spherical detector, consistent with the injected activated brine having a higher density. Once the circulation was turned on, the heavy water was pumped out of the bottom and returned near the top of the acrylic vessel. As a result, the activity moved from the bottom to the top, around the outside and, eventually, into the middle of the vessel.

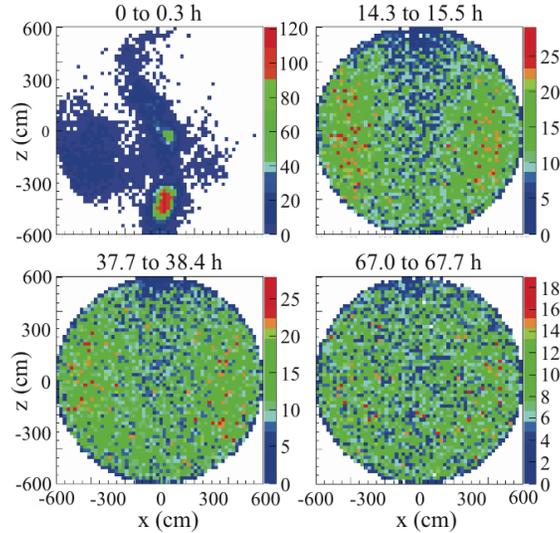

Figure 7: Location of $^{24}$Na activity inside the acrylic vessel, following spike ♯5, along a vertical slice enclosing the centre of the vessel. The panels correspond to different times after injection of the spike in October 2005. The spike was fully mixed after 68 h. The colour indicates the activity in the bin (counts/time).

A faster way to mix the sources was found later for the final $^{222}$Rn and $^{24}$Na spikes (♯6 and ♯7) injected in the heavy water in August and October 2006, respectively, in phase III. The method used flow reversal and temperature inversion to create extensive eddy current mixing. Analysis of the movement of earlier spikes showed that the $D_2O$ circulation caused a rotation of the entire bulk of $D_2O$ over a period of time. This was not unexpected as a flow into a large tank generally (because of slight asymmetries) provides rotational momentum and slowly builds up rotational flow of some kind.

About ten hours prior to the spike, the $D_2O$ circulation was started in the AV in the normal flow direction, where water is removed from the bottom and returned to the top of the AV. This set up some rotational motion of the $D_2O$. Just before the spike was introduced directly into the water flow going to the AV, the direction of the water flow was reversed; *i.e.* removed from the top and returned to the bottom. This caused eddy currents and mixing. In addition, the chilled water flow to the heat-exchanger was stopped, so that the water returned to the bottom $1 - 3$ C warmer than the bulk of the $D_2O$, resulting in the returned water rising and more thoroughly mixing throughout the bulk.

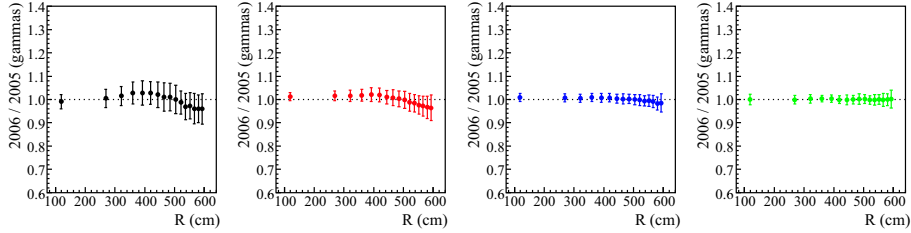

Figure 8: Mixing of the $^{24}$Na: the ratio of the 2006 and 2005 radial event distributions. Time increases from left to right and the rightmost plots correspond to the steady state, which was attained after 68 h in 2005 and 51 h in 2006.

Also, any water accumulating at the AV top would get circulated back to the bottom. With this process the $^{24}$Na was fully mixed 51 h after the spike was added.

For the final $^{222}$Rn spike ($\sharp$8) at the end, the radon was injected into the inner light water shield (see Fig. 1) at the end of phase III via the N$_2$ re-gassers of the main circulation loop. The injected water temperature was first adjusted to be as close as possible to the ambient temperature. After about two hours, the temperature was alternately increased and decreased every two hours in order to try and mix the injected radon. The water in the inner shield was also circulated to help with the mixing.

## 4. Results from distributed $^{24}$Na and $^{222}$Rn sources

Four measurements will now be discussed that illustrate the importance of the distributed sources of $^{24}$Na and $^{222}$Rn: 1) the neutron detection efficiency of the NCD array using two uniformly mixed sources of $^{24}$Na; 2) the neutron response of the NCD array; 3) the calibration of the in situ background analysis technique using $^{24}$Na; and 4) the response of the SNO detector to $^{222}$Rn.

### 4.1. The neutron efficiency of the NCD array

When using a mixed $^{24}$Na spike to measure the neutron efficiency of the NCD array, a measure of the uniformity of the distribution of $^{24}$Na in the heavy water was required. The uniformity was assessed using the combined light from the resulting beta decay electron (end-point energy 1.393 MeV) and the 1.369 MeV and 2.754 MeV $\gamma$-rays, which gave a spectrum of energy deposition peaking at an effective kinetic energy of $\sim$3.6 MeV.

The detector was divided into radial shells of equal volume and the number of events reconstructing in each shell was determined. While the detection efficiencies changed with radius by as much as $\sim$20%, a spike was judged to be fully mixed (the steady state) when the standard deviation of the number of events in the detector volume elements was no longer changing within the statistical uncertainties.

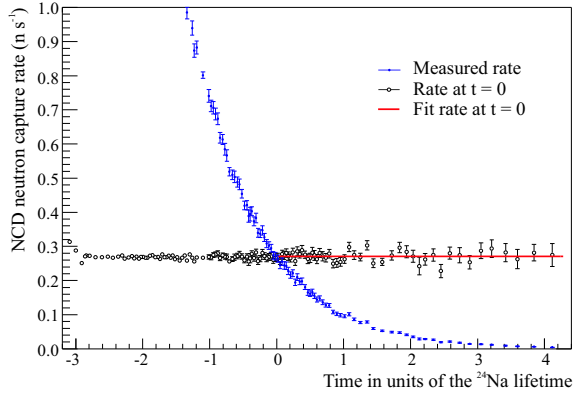

Figure 9: The neutron capture rate of the 2005 $^{24}$Na spike as a function of time, showing the data, the rate corrected to the time $t = 0$ that the source was fully mixed, and the fitted rate at that time. $t = 0$ was 4.53 lifetimes after the spike was added.

Evidence that spikes were then uniformly mixed in regions away from the thin boundary layers (i.e. in the bulk of the heavy water), besides that obtained from studying the water flow of individual spikes, came from a comparison of the 2005 and 2006 spikes. In each of these the source was injected very differently, as described above in Section 3.2, and the subsequent mixing set up distinct currents of $^{24}$Na in the heavy water. As shown in Fig. 8, the final event distributions in these measurements were essentially identical. It is improbable that the $^{24}$Na could converge to the same spatial distribution in 2005 and 2006, and it not be a uniform one. Furthermore, neither distribution would be static.

This conclusion was also supported by the good agreement of the neutron capture rate with an exponential decay (see Fig. 9). Furthermore, a comparison of the distribution before complete mixing to when fully mixed showed that the standard deviation of the ratio of concentrations for each region decreased as the water mixed and reached the statistical limit. This comparison limited the residual concentration differences in the bulk of the heavy water to less than 1%.

A study of the Cherenkov light yield from the mixed $^{24}$Na source near the AV showed that there was no unmixed layer of $^{24}$Na next to the AV with an uncertainty of $\sim 3$ cm, consistent with uniform mixing. The dimensional estimate for the thickness of the NCD boundary layer is $\sim 5$ centimetres.

Including the total uncertainty from non-uniformity (1.6%), the error on the rates of the spikes (1.2%) and that on the instrumental efficiency (1.2%), the neutron detection efficiency for the NCD array was determined to be $0.211 \pm 0.005$ for the NC signal. This value was in excellent agreement with the Monte Carlo prediction of $0.210 \pm 0.003$. Details of these calculations will be provided in a future publication.

*4.2. The neutron response of the NCD array*

The NC neutron response to the $^3$He proportional counter array in phase III of SNO was measured with the mixed $^{24}$Na source, as the neutron captures are distributed over the array geometry in the way expected for the NC signal. The energy spectrum obtained is shown in Fig. 10. The capture of a neutron by $^3$He results in a proton and a triton particle with kinetic energies of 573 keV and 191 keV, respectively. The indicated edges in the spectrum correspond to just detecting either a proton or a triton particle. This spectrum was used as a probability density function (pdf) in the extraction of the NC flux in phase III of SNO [7].

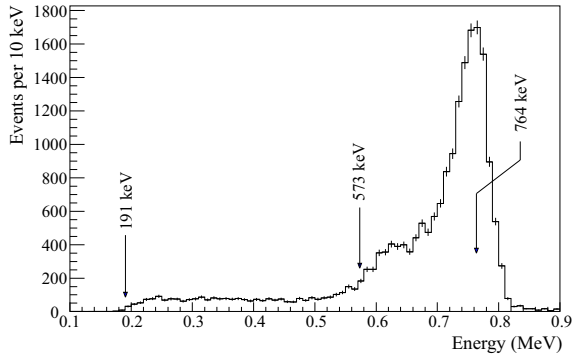

Figure 10: $^{24}$Na NCD energy spectrum measured with spike ♯5. Data shown is after data cleaning cuts have been applied.

*4.3. The detector response to the $^{24}$Na and $^{222}$Rn spikes in phase II*

For the accurate determination of the CC and NC fluxes, it was very important to determine all backgrounds precisely. In particular, there was an indistinguishable background to the Neutral Current signal from trace amounts of natural radioactivity. The $\beta - \gamma$ decays of $^{208}$Tl and $^{214}$Bi from the naturally occurring $^{232}$Th and $^{238}$U chains emit $\gamma$-rays that are above the deuteron binding energy. The neutrons produced through photodisintegration were indistinguishable from those produced by the NC reaction and, therefore, the measurement of the levels of radioactivity inside the detector was crucial for a correct measurement of the total $^8$B flux.

One way of determining the amounts of $^{208}$Tl and $^{214}$Bi, as discussed in section 1.2, was based on an analysis of the number of events that occurred in a low energy window dominated by the $\beta - \gamma$ decays from these nuclei. The method, called the in situ technique, relied on using pattern recognition on the Cherenkov light distribution to determine their relative concentrations. The light distribution was characterized by the parameter $\beta_{14} \equiv \beta_1 + 4\beta_4$, where $\beta_l$ was the average value of the Legendre polynomial $P_l$ of the cosine of the angle between PMT hits.

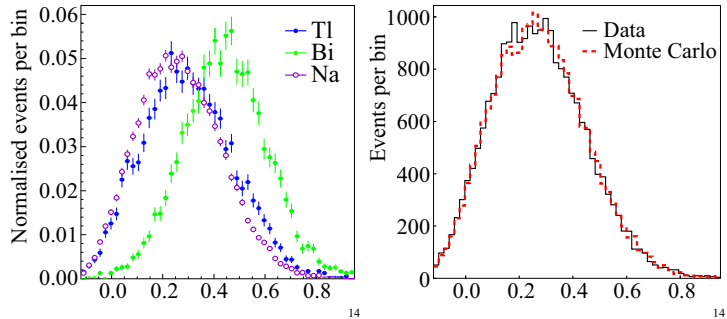

Figure 11: A comparison of the $\beta_{14}$ distributions: $^{208}$Tl, $^{214}$Bi and $^{24}$Na on the left; $^{24}$Na data and Monte Carlo simulation on the right.

For $^{208}$Tl, the Cherenkov light in this energy window, which was between $4.0 - 4.5$ MeV in terms of effective kinetic energy, is from a $\beta$ and at least one $\gamma$-ray while, for $^{214}$Bi, it is dominated by that from the $\beta$ decay to the ground state of $^{214}$Po. As a result, $^{208}$Tl produces a more isotropic light distribution than that from $^{214}$Bi and has a correspondingly smaller $\beta_{14}$. This difference, as shown in the left plot in Fig. 11, allowed $^{208}$Tl and $^{214}$Bi to be separated on a statistical basis [4].

To determine the response to $^{208}$Tl and $^{214}$Bi decays, uncontained sources of $^{24}$Na and $^{222}$Rn were used to avoid the effect of attenuation of the betas in contained sources. The $\beta_{14}$ distribution of $^{24}$Na is similar to that of $^{208}$Tl (see Fig. 11) as its Cherenkov light is also from a $\beta$ and $\gamma$-rays, while $^{222}$Rn decays to $^{214}$Bi, the subsequent decay of which is observed. Therefore, the distributed sources $^{24}$Na and $^{222}$Rn provided calibrations of the $^{208}$Tl and $^{214}$Bi $\beta_{14}$ distributions.

As shown in Fig. 11, good agreement was found between data from the first central $^{24}$Na spike in phase II (Table 1, ♯1), described in Section 2.3, and the Monte Carlo simulation for the angular distribution of Cherenkov light as measured by $\beta_{14}$. For the region within which the source was mainly contained, $r \leq 230$, the $\beta_{14}$ means were: data $0.2472 \pm 0.0031$; Monte Carlo $0.2484 \pm 0.0036$.

Initially, a discrepancy of approximately 2.5% was observed in $\beta_{14}$ distributions. However, good agreement was obtained after a small correction was made to EGS4 [18] to account for approximations used in the description of electron scattering in the MC simulation, mainly for the neglect of the Mott terms [4].

The observed isotropy distribution from the mixed Rn-dataset from spike ♯3 was compared to Monte Carlo simulation and these were also found to be consistent [23]. In addition these $^{24}$Na and $^{222}$Rn spikes were used to check the energy scale at these low energies, which was important in calibrating the in situ technique.

## 4.4. Measurement of the $^{222}$Rn energy spectrum and induced neutron background rate

A radon spike in the heavy water in phase II allowed the measurement of the dominant component of the radioactive background below 4.5 MeV effective kinetic energy. Radon spike data also gave a direct measure of the ratio of neutron to Cherenkov events and valuable cross-checks on the energy scale and resolution at low energies as well as on photodisintegration cross-sections and neutron capture efficiencies.

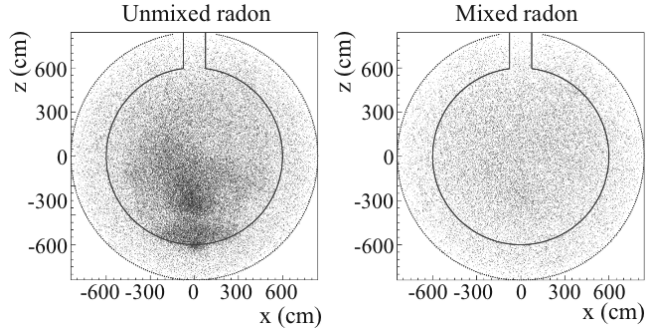

Figure 12: Vertical scatterplot for the unmixed and mixed radon spike data set obtained with spike ♯3.

### 4.4.1. The $^{222}$Rn energy spectrum

Radon assays taken during phase II from different positions in the acrylic vessel indicated that the radon concentration might have been up to three times higher near the neck than at the bottom.

In the radon spike ♯3 in phase II, described in Section 2.1 above, the spike was left unmixed for four days and then the circulation was turned on and the radon was distributed more uniformly throughout the $D_2O$ as can be seen in Fig. 12. The events seen outside the AV are mainly from mis-reconstructed events and radon in the $H_2O$.

The energy spectra from the unmixed and mixed data sets agreed as well as could be determined by their statistics. By using both data sets the errors on the $^{222}$Rn energy spectrum arising from uncertainties in the actual $^{222}$Rn distribution could be estimated.

The resulting $^{222}$Rn energy spectrum from the mixed data set is shown in Fig. 13. The data were fit using the FTK energy estimator [24] and using a Kernel estimator [25]. The energy resolutions between the Monte-Carlo and data agree to within 0.5% at 4.5 MeV [26].

### 4.4.2. Measurement of the external background from the $H_2O$

Radioactivity in the PMTs, the PMT support structure and in the light water shield could give rise to backgound events in the $D_2O$. Some $\gamma$-rays could penetrate the AV or be mis-reconstructed in the $D_2O$, and those with energy greater than 2.225 MeV could photodisintegrate deuterons, thus forming a neutron background. These backgrounds were strongest near the edge of the vessel and quickly dropped off for smaller radii.

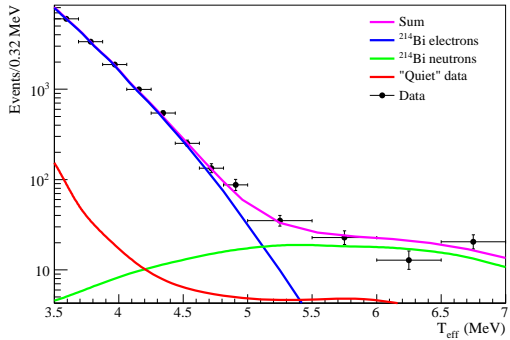

Figure 13: Energy spectrum of the radon spike data obtained with spike ♯3 and the fitted MC prediction, showing the steeply-falling $\beta-\gamma$ Cherenkov events (due to progeny $^{214}$Bi) and the neutron peak.

Contained sources of $^{232}$Th and $^{238}$U were used to measure the response to these $\gamma$-rays. However, to calibrate the effect of mis-reconstructed $\beta-\gamma$ decays as well as from $\gamma$-rays from activity in the light water, an uncontained source of $^{222}$Rn was used so that the response to $\beta-\gamma$ decays could be measured.

To study these backgrounds a radon spike of $50 \pm 2.4$ kBq (see Section 2.2 above) was injected into the light water shield at the end of phase III via the $N_2$ re-gassers of the main circulation loop. After injection and mixing of the radon in the light water, the starting neutron-rate, as observed in the NCD array, was found to be 1.82 neutrons per hour [27]. Comparing the injected activity to the activity in the light water shield, the total number of neutrons expected to be detected in the NCD array during the whole of phase III was ∼two neutrons from the light water shield, to be compared with the ∼983 induced by solar neutrinos in phase III.

## 5. Conclusions

Distributed uncontained sources of $^{24}$Na and $^{222}$Rn were used in SNO to provide important calibrations of the response of the SNO detector to neutrons and to low energy $\beta-\gamma$ decays from traces of natural radioactivity within the detector. A uniformly mixed source of $^{24}$Na proved particularly useful in determining the neutron detection efficiency of the array of $^3$He proportional counters deployed in the last phase of SNO, while a dispersed $^{222}$Rn source provided valuable information on the low energy backgrounds, both internal ($D_2O$) and external ($H_2O$). Both sources provided useful calibrations of the response to Cherenkov light from low energy $\beta - \gamma$ decays, important in determining the number of background neutrons produced through photodisintegration of deuterons in the heavy water by trace amounts of $^{208}$Tl and $^{214}$Bi.

The techniques for producing sources of $^{24}$Na and $^{222}$Rn described in this paper could be useful for the calibration of future low background detectors.

## 6. Acknowledgements

The authors are very grateful to the SNO collaboration, the site operations crew and to Vale INCO and their staff at Creighton mine, without whose help this work could not have been conducted, and would like to thank Atomic Energy of Canada, Ltd. (AECL) for the generous loan of the heavy water in cooperation with Ontario Power Generation. They would also like to thank the Royal Military College, Kingston, Ontario, for their help with the production of $^{24}$Na.

This work was supported in the United Kingdom by the Science and Technology Facilities Council (formerly the Particle Physics and Astronomy Research Council); in Canada by the Natural Sciences and Engineering Research Council, the National Research Council, Industry Canada, the Northern Ontario Heritage Fund Corporation, and the Province of Ontario; and in the USA by the Department of Energy. Further support was provided by Vale INCO, AECL, Agra-Monenco, Canatom, the Canadian Microelectronics Corporation, AT&T Microelectronics, Northern Telecom, and British Nuclear Fuels, Ltd.